\documentclass[aps,twocolumn,groupedaddress,prb]{revtex4}
\usepackage{graphicx}%

\begin{document}
\bibliographystyle{apsrev}


\title{Heterogeneity, and the secret of the sea}


\author{P. Fraundorf}
\affiliation{Physics \& Astronomy, U. Missouri-StL (63121), St. Louis, MO, USA}


\date{\today}

\begin{abstract}

	This paper explores tools for modeling and 
measuring the compositional heterogeneity of a 
rock, or other solid specimen.  Intuitive 
``variation per decade'' plots, simple expressions 
for {\em containment probability}, generalization 
for familiar {\em error-in-the-mean} expressions, 
and a useful dimensionless {\em sample bias coefficient} 
all emerge from the analysis.  These calculations 
have also inspired subsequent work on log-log 
roughness spectroscopy (with applications to 
scanning probe microscope data), and on angular correlation 
mapping of lattice fringe images (with applications 
in high resolution transmission electron 
microscopy).  It was originally published as 
Appendix E of a dissertation\cite{Fraundorf80} on ``Microcharacterization 
of interplanetary dust collected in the earth's stratosphere''.
\end{abstract}
\maketitle

\tableofcontents

\section{2nd moment statistics: next after mean}
\label{sec: secondmoment}

An intuitively simple, but quantitative, approach to spatial heterogeneity 
is possible using mathematics which claims wide familiarity because of its 
already diverse applications.  The composition of a specimen can be 
considered a function of spatial position whose statistical properties 
are fully described by a set of underlying probability distributions\cite{Wang45}.  
These distributions are fully specified 
by knowledge of their moments.  Naturally, the first parameters to measure 
in describing these distributions are the first moments: the average 
values or compositional abundances.  Elemental and mineral (modal) abundances 
have long been familiar tools in the characterization of geological specimens.

The logical next step in specifying the distributions is to use information 
on point-to-point variations to measure the second moments of the probability 
distributions.  If, for simplicity and in the interest of improved 
sample statistics, the underlying distributions are considered invariant 
under translations and rotations, then the statistics of point-to-point 
variations are fully described by the set of one-parameter functions known 
as radial covariances.  If these functions are measured (using a three-dimensional 
array of sample points) for a specimen whose distributions are not 
invariant under rotations and translations (e.g. in samples with 
aligned crystals or a compositional gradient), then these functions will 
represent averages (over direction and position) of covariance functions 
which depend on orientation and absolute position in the sample.  Of course, 
if one wishes to {\em infer} statistical properties of the parent material 
from a laboratory sampling, it is necessary (and traditional) to assume 
translational invariance in the absence of contrary information.  
Because inferences from samples volume to the larger specimen are 
an important aspect here, a definition of source material extending only 
to material in which the underlying distributions are ``stationary'' will 
be implicit in the discussion to follow.  Rotational invariance (or 
isotropy) for the underlying distributions will not be taken for granted, 
although its assumption will be necessary if one wishes to measure the 
radial covariances from polished sections which do not sample a wide 
variety of orientations in the specimen.

Applications for the concept described above are most familiar in 
the field of time series analysis\cite{Blackman59,Bloomfield76}, 
where the one-dimensional independent variable, 
time, is used in place of the three-dimensional variable, position.  
However, the concept itself is quite general, and myriad two and 
three dimensional spatial applications already exist in the 
literature\cite{Wu72,Peebles73}.  In fact, curiousity 
about the statistical properties of composition in rocks was inspired 
with an article by Martin Gardner\cite{Gardner78} concerning applications 
to natural phenomena ranging from the topology of cratered 
terrain\cite{Mandelbrot77} to the pattern of notes and rhythm in a 
piece of music\cite{Voss75}.

The general covariance function (or cross-covariance) of variables 
$A$ and $B$ evaluated at regions displaced by $\Delta \vec x$ is defined 
by 
\begin{equation}
\sigma _{AB} [\Delta \vec x] \equiv \langle \{ A[\vec x + \Delta \vec x] - \mu _A \} \{ B[\vec x] - \mu _B \} \rangle _{\vec x}. 
\label{gencovariance}
\end{equation}
Conventional notation will be used here:  population (or ``parent specimen'') averages 
are denoted by angle brackets $\langle \rangle$ or Greek letters, while sample 
(i.e. measured) averages are denoted by a bar over the quantity, or Roman letters.  
In particular, $\mu_A$ and $\mu_B$ are defined as the population averages of $A$ 
and $B$ respectively, with the averages here taken over all possible values of
location vector $\vec x$.  {\em Note}:  This quantity is sometimes put into 
dimensionless (correlation coefficient) form by normalizing with standard 
deviations of $A$ and $B$, as in
$\rho_{AB}[\Delta x] \equiv \frac{\sigma_{AB}[\Delta x]}{\sigma_A \sigma_B}$.

For the petrographic application, it is useful to examine a mixture 
of $n$ homogeneous phases.  Consider spatial heterogeity for the variables 
$A$ and $B$ which take on values of ${}_iA$ and ${}_iB$, respectively, 
in the ith phase, $i=1,n$.  If $p[i,j,t]$ is the {\em joint} probability 
that, for two sample points separated by a distance (or lag) $t$, the 
first will lie on a phase $i$ region and the second will lie on a phase 
$j$ region, then the radial covariance function for $A$ and $B$ can be 
written:
\begin{equation}
\sigma _{AB} [t] = \sum\limits_{i = 1}^n {\sum\limits_{j = 1}^n {p[i,j,t]({}_iA - \mu _A )({}_jB - \mu _B )} }. 
\label{probsum}
\end{equation}
If $A \equiv B$, then the quantity defined above is referred to as the radial 
auto-covariance for $A$.

The $\sigma _{AB} [t]$ constitute a measure of the statistical 
correlation between the value of the parameters $A$ and $B$ at 
points in the specimen separated by a distance $t$.  For the special 
case $t=0$, the radial covariance $\sigma _{AB} [t]$ reduces to 
the covariance for $A$ and $B$ in the specimen, i.e. $\sigma _{AB} [0] = \sigma _{AB}$. 
Similarly the radial auto-covariance for $A$, $\sigma _{AA} [t]$, reduces to 
the variance of $A$ in the specimen, i.e. $\sigma _{AA} [0] = {\sigma_{A}}^2$. 
Using this notation, the absolute value of $\sigma _{AB} [t]$  for the specimen 
is less than or equal to the product of standard deviations 
$\sigma_A \sigma_B$, and it is equal to zero when 
values of $A$ and $B$ measured at points separated by length $t$ are 
uncorrelated.

Although the general covariance functions contain all information on 
the statistics of point-to-point variations in a specimen, the value of 
the function at any chosen argument depends upon heterogeneities that 
may have many different length scales\cite{Peebles73}.  Fortunately, 
knowledge of covariance functions can be converted, without loss of 
detail, to knowledge of spectral densities which serve to separate 
effects due to heterogeneity on various size scales.  In this way, 
data obtained by techniques which are sensitive to heterogeneity on 
different size scales can be combined to provide a picture of 
compositional variability over a wide range of spatial frequencies.  

If we mimic equation \ref{gencovariance} and 
define {\em correlation} $C_{AB}[\Delta \vec x]$ as the average over all 
$\vec x$ of $A$ times $B$ at regions displaced by $\Delta \vec x$, one 
can quite generally write $\sigma_{AB} = C_{AB} - \mu_A \mu_B$, where 
the product of the means is a constant with no dependence on 
$\Delta \vec x$.  But the Fourier correlation theorem\cite{Press92} says that 
correlation is the inverse transform of the conjugate product of spectral 
densities for $A$ and $B$ in the specimen, and $\mu_A$ and $\mu_B$ 
are volume-normalized spectral densities measured at zero frequency.  
Thus the Fourier transform of $\sigma_{AB}$ is the conjugate 
product of volume-normalized spectral densities for $A$ and $B$, with 
the zero-frequency value set to zero.  

To be specific, equation \ref{probsum} defines a function of the magnitude 
of the separation between two points in the sample, but not of the 
direction of that separation.  Its three-dimensional Fourier transform (using 
signal processing conventions) can be written as:
\begin{equation}
\gamma _{AB} [f] \equiv 2 \int\limits_0^\infty {\frac{t}{f}} \sin [2\pi ft] C_{AB} [t]dt.
\label{forwardtransform}
\end{equation}
The inverse equation for this expression is:
\begin{equation}
C_{AB} [t] \equiv 2\int\limits_0^\infty  {\frac{f}{t}} \sin [2\pi ft]\gamma _{AB} [f]df.
\label{inversetransform}
\end{equation}
For the case $t=0$, this yields the relationships:
\begin{equation}
C_{AB} [0] \equiv \int\limits_0^\infty 4\pi {f^2} \gamma _{AB} df = \int\limits_{ - \infty }^\infty  {4\pi f^3 } \gamma_{AB} d(\ln f).
\label{t0case}
\end{equation}
Hence one might relate spectral density physically to $4\pi {f^2} \gamma _{AB} $, 
which from above is that portion of the covariance 
$\sigma_{AB}[0] = C_{AB}[0]-\mu_A \mu_B$ resulting
from compositional fluctuations occurring with spatial frequency in the 
interval from $f$ to $f + df$.  For a quantity independent of the units used to 
measure frequency, the second equality in \ref{t0case} relates spectral 
density to $4\pi {f^3} \gamma _{AB} $, that portion of the covariance between $A$ 
and $B$ in a given ``e-fold'' of frequency (or decade of frequency should one 
further multiply by $\ln 10$).

One example is the special case when all variations 
occur with one frequency, $f_o$.  Sinusoidal banding of an 
otherwise homogeneous material would, for instance, give rise 
to this condition.  Regardless of details of the underlying 
distribution, however, the radial auto-covariance and 
spectral-density ({\em sans} origin) for a variable of this type can be written:
\begin{equation}
\sigma _{AA} [t] \equiv \sigma _A ^2 \frac{{\sin [2\pi f_o t]}}{{2\pi f_o t}},
\label{sinesigma}
\end{equation}
and
\begin{equation}
\gamma _{AA} [f] \equiv \sigma _A ^2 \frac{{\delta [f - f_o ]}}{{4\pi f_o ^2 }}.
\label{sinegamma}
\end{equation}

\section{The well-mixed model}
\label{wellmixed}

Although long-range order, like that described by equations \ref{sinesigma} 
and \ref{sinegamma}, is certainly possible in geological materials 
(e.g. in compositional zoning of mineral single crystals), it is likely 
to be the exception rather than the rule in most aggregate materials.  
A more likely model is one which assumes that the specimen consists of 
single-phase regions whose relative positions are fully random (that is, 
well-mixed) with respect to one another.  If the regions are atomic 
dimensions in size, then the model describes a glass; if the single-phase 
regions are considerably larger, then the model would describe, for example, 
a well-mixed breccia.  In igneous or metamorphosed materials, the 
possibility of compositional correlations between adjacent crystals could 
give rise to deviations from this well-mixed case.

Even a well-mixed model could be quite difficult analytically if one 
wishes to consider the details of the interplay (in three dimensions) 
between grain shapes, sizes, and the packing arrangement.  However, 
considerable simplification results if one assumes short range order in 
the sense that lage, which leave their initial grain, necessarily 
terminate on points compositionally uncorrelated with the starting 
location.  If we define $p[i]$ as the fractional abundance of phase $i$ 
points, and $\xi_i[t]$ as the probability that lag $t$, beginning at 
an arbitrary location on a grain of type $i$, will {\em not} leave 
that grain before ending, then we can write:
\begin{equation}
\sigma _{AB} [t] = \sum\limits_{i = 1}^n {p[i]({}_iA - \mu _A )} ({}_iB - \mu _B )\xi _i [t],
\label{mixedgraincov}
\end{equation}
and
\begin{equation}
\gamma _{AB} [f] = \sum\limits_{i = 1}^n {p[i]({}_iA - \mu _A )} ({}_iB - \mu _B )\Xi _i [f].
\label{mixedgrainspden}
\end{equation}
The function $\Xi_i [f]$ is the radial Fourier transform of the 
{\em containment probability} $\xi_i [t]$, which in turn obeys the 
simple constraint:
\begin{equation}
4\pi \int\limits_0^\infty  {\xi _i [t]t^2 dt = \Xi _i [0] = \langle V_i } \rangle ,i = 1,n.
\label{volumeconstraint}
\end{equation}
Here, $\langle V_i \rangle$ is the (volume-weighted) average 
volume for grains of type $i$.

Calculations for several simple grain geometries have been attempted.  
For spherical grains of diameter $D$, the containment probability 
is\cite{Plachy80}:
\begin{equation}
\xi _i [t] = \{ \begin{array}{*{20}c}
   {1 - \frac{3}{2} \left({\frac{t}{D}}\right) + \frac{1}{2} \left({\frac{t}{D}}\right)^3 } & {, t \le D.}  \\
   0 & {, t > D.}  \\
\end{array}
\label{sphericalcase}
\end{equation}
Of course, not only are truly spherical grains rare, but a 
model made up solely of spheres of one size necessarily is 
filled with voids.  

A more realistic model for containment probabilities comes 
from the case of rectangular solids.  Unfortunately, a full 
analytical solution has not yet been obtained.  The case 
for a cube of side $d$ has been tabulated by a combination 
of analytical and numerical methods.  Somewhat by accident 
it was discovered that an RMS deviation from the exact 
values of only 0.0045, over the non-zero range $0 < t < \sqrt3 d$, 
is obtained with the much simpler approximation:
\begin{equation}
\xi _i [t] \cong \{ \begin{array}{*{20}c}
   {\left(1 - {\frac{t}{{bd}}} \right)^2 } & {, t \le bd.}  \\
   0 & {, t > bd.}  \\
\end{array}
\label{cubicapprox}
\end{equation}
Here $b$, from the volume constraint \ref{volumeconstraint}, 
must equal $\left({\frac{30}{4 \pi}}\right)^{1/3}$.  The Fourier 
transform of the approximation is also similar to the exact 
one, although the low level ``ripples'' for values of $f$ 
large compared to $1/d$ are differently placed.  These ripples 
are of course averaged away when a continuum of grain sizes 
are used, and hence should have no effects in applications 
to real materials.  Thus the approximation given in \ref{cubicapprox} 
for a cube of side $d$ will be adopted as a simple expression for 
work here.

The equation \ref{mixedgraincov} provides a model for mixed-grain 
radial covariances, given the phase abundances $p[i]$ and the containment 
probability function $\xi_i [t]$ for grains of each phase.  Note that 
in this well-mixed case, auto-covariance will always be a positive 
monotone-decreasing function of lag which goes to zero for lags large 
compared to ``compositional'' grain sizes in the specimen.  In particular, 
if the grain geometry (in terms of the containment probability 
function) is the same for all phases, then all of the $\sigma_{AB} [t]$ 
for the specimen exhibit the same $t$ dependence.  For such a specimen, 
knowledge of the $t$ dependence for the radial auto-covariance of one 
variable would specify the $t$ dependence for all.  Conversely, if the 
$n$ phases obey one of $m$ different containment probability 
functions ($m \le n$), then measurement of the radial auto-covariance 
for $m$ variables might allow calculation of the containment probabilities 
for each phase type via \ref{mixedgraincov}, as well as a check of the 
model by predicting values of the radial cross-covariances (i.e. 
cases in which $A \ne B$).

Given that many geological samples can be considered neither 
homogeneous glasses nor well-mixed breccias, then what deviations from 
the well-mixed model are to be expected?  One simple deviation can 
be detected by checking the relationship between physical grain size 
and compositional grain size.  If the two do not agree, then a more 
complicated history is required.  For example, a physical grain size 
much smaller than the compositional grain size might result from a 
breccia composed of larger fragments which were crushed by not well-mixed 
prior to compaction.  Conventional petrographic analysis, 
unsurpassed in its ability for qualitative discernment of patterns 
requiring knowledge of higher statistical moments, would be helpful 
in verifying the cause even if the effect itself required the 
quantitative approach described here.

In a wider variety of ways, well-mixed materials can be 
modified by open and closed-system thermodynamic processes.  
If the scale of transport allowed in such processes is large 
compared to the sample size, especially in the presence of 
a long-range gravitational force, then the resulting effects on 
specimen heterogeneity must be handled case by case.  But if 
in-situ transport processes are limited to distances smaller 
than the specimen size, then one overall effect of the redistribution 
of material is predictable:  regions enriched on one element 
will be surrounded by regions depleted in that element.  
This means that the probability of finding such a redistributed 
element is likely to decrease upon leaving an enriched region, 
prior to leveling off at the average probability.  Hence 
the radial auto-covariance for that element is likely to go 
below zero before it dies out as $t$ gets larger.  
In the limiting case when an element-pure region (phase 1) 
is small compared to a surrounding element-free region 
(phase 2), then the negative excursion for $\sigma_{AA} [t]$ 
goes down to:
\begin{equation}
min[\sigma_{AA} [t]] \cong - \left( \frac{b_1}{b_2} \right)^3 \sigma_A^2 .
\end{equation}
Here $b_1$ is a characteristic dimension of the enriched zone and $b_2$ 
a typical size for the depleted zone.

The possibly diagnostic nature of such deviations from 
the well-mixed case suggests that measurement of these 
quantities from actual samples, in addition to modeling 
on the basis of the observation of phase abundances 
and geometries, might be a worthwhile pursuit.  This is 
especially true in light of the fact that information 
on point-to-point compositions in geological materials 
is frequently available, and the data on relative 
locations of the sampled points easy to obtain, during 
the course of automated modal analyses.

\section{Measurement from compositional data}

Compositional information, as a function of position in a 
specimen, is most conveniently measured on flat sections of 
the specimen.  This is true whether compositional identifications 
are made by optical microscopy, polished-section energy or 
wavelength dispersive x-ray analysis, or thin-specimen analytical 
transmission electron microscopy.  As mentioned at the outset, 
measurement of radial covariances from ``two-dimensional'' samples 
of this sort requires either (1) an assumption that the specimen 
analyzed is isotropic, or (2) averaging of the data for sections 
which sample a wide range of orientations in the specimen.  
A third alternative is to use slices through the specimen taken 
in a why such that the relative three-dimensional positions 
of the sampled locations are known.  Observations of the 
chondritic interplanetary dust aggregates examined in this 
thesis, with SEM and TEM, provide no indication that the 
assumption of radial isotropy is unwarranted at this point.

As will be easier to see in examples of the Fourier transformed 
radial covariances, the spatial resolution of the sampling 
method and the maximum separation between points considered in 
the covariance calculation determine, respectively, the upper 
and lower bounds (in terms of spatial frequency) on the ``spectral 
window'' through which specimen heterogeneity is being examined.  
In order to avoid aliasing it is useful to set the minimum 
spacing between analysis points (the ``lag interval'') to be on the 
order of (or smaller than) the spatial resolution of the point 
analyses.  In order to improve spectral resolution and increase 
the number of effectively independent sample locations, it is important 
to sample as wide a range of lags (point separations) as is possible. 
In meeting these criteria for a fixed number of point analyses, 
a simple square array of sampled locations is not the optimum 
configuration, but it is one which will be frequently encountered. 
In the discussion to follow, we will assume quite generally an 
array of $M$ sample locations: $A_i$ and $B_i$ are the values 
measured for $A$ and $B$ at the ith sample location (position $\vec x_i, i=1,M$).
Thus the postscripts on $A$ and $B$ refer to sampled locations, not 
to mineral phases.

Although a number of slightly different ``sample'' cross-covariance 
functions could be defined, it is simplest in the interest of subsequent 
error analysis to use for $\mu_A$ and $\mu_B$ the best available 
estimates for the mean values of $A$ and $B$ in the specimen, 
and then to calculate:

\begin{equation}
s_{AB} [t] \equiv \left( {\frac{1}{M_t}} \right)\sum\limits_{i = 1}^M {\sum\limits_{j = 1}^M {\delta _{tij} (A_i  - \mu _A )} } (B_j  - \mu _B ).
\label{samplecov}
\end{equation}
Here $\delta _{tij}$ is simply a delta function used to disallow from 
the double sum all terms except those for which analysis points $i$ and 
$j$ are separated by a distance $t$ in the specimen, and $M_t$ is just the 
double sum over that delta function (i.e. the number of point pairs 
in the sample separated by a distance $t$).  Of course, any two or three 
dimensional sample configuration is likely to include pairs which are 
not all integral multiples of the (minimum) lag interval, $\Delta t$.  
In order to create a set of equally spaced data points for the Fourier 
transform process, it is useful to ``widen'' $\delta_{tij}$ to allow 
all pairs into the sums for $t$ which have separations within $\frac{1}{2} \Delta t$ 
on either side of $t$, and then to evaluate \ref{samplecov} for all 
$t_j = (j-\frac{1}{2}) \Delta t, j=1,T/\Delta t$.  Here $T$ is the 
maximum lag introduced into the calculation.  Because of the large 
uncertainties associated with data on lags larger than half of the 
sample size, $T$ will usually be chosen to be less than half of the maximum 
separation between points in the sample.

As with any estimate of the radial covariance based on 
measurements from a finite sample, \ref{samplecov} provides a 
biased estimate.  If we define restricted averages, based only 
on data pairs separated by the distance $t$, of the form:
\begin{equation}
\bar A_t  \equiv \left( {\frac{1}{{M_t }}} \right)\sum\limits_{i = 1}^M {\sum\limits_{j = 1}^M {\delta _{tij} A_i } }, 
\end{equation}
then with simple algebra it is easy to verify that:
\begin{equation}
\langle s_{AB} [t]\rangle  = \sigma _{AB} [t] - \langle (\bar A_t  - \mu _A )(\bar B_t  - \mu _B )\rangle .
\label{sampop}
\end{equation}
If lags of all sizes are uniformly distributed over the 
sample, then the second term in \ref{sampop} will depend very little 
on $t$, and be approximately equal to $\langle (\bar A  - \mu _A )(\bar B  - \mu _B )\rangle$, the covariability in estimates of $\mu_A$ and $\mu_B$ from the sample.  
Thus when $A \equiv B$, the probable bias in the estimate \ref{samplecov} is 
just the variance in the estimate of $\mu_A$, which cannot be determined 
without assumptions about the character of the parent specimen (see next 
section).  The amount of bias is roughly the same for all $t$, and hence 
\ref{samplecov} provides an unbiased estimate for the shape of 
$\sigma_{AB} [t]$, but not for its absolute position in the vertical 
direction.  On the other hand, the bias in estimates of the spectral 
density is not independent of frequency.  Before estimation of the 
spectral density, however, the abrupt cutoff in the radial covariance 
estimate at $t=T$ must be addressed.

In order to avoid forcing the calculated covariance function 
$s_{AB}[t]$ abruptly to zero for $t \ge T$, it is 
traditional\cite{Blackmank59} to do this gradually prior to 
Fourier transformation by multiplying the measured covariance 
function by a function which goes ``gently'' from 1 down to zero 
over the interval $0<t<T$.  A common choice of this function 
is the Hanning lag window, defined by:
\begin{equation}
D[t] = \{ \begin{array}{*{20}c}
   {{\textstyle{1 \over 2}}(1 + \cos [{\textstyle{{\pi t} \over T}}])} & {,t \le T}  \\
   0 & {,t > T}  \\
\end{array}
\end{equation}
The finite range of lags allowed into the calculation for $S_{AB}[t]$ 
inevitably decreases the resolution of the measured spectral density.  
Gentle supression of the covariance function for large lags prior 
to Fourier transformation results in a simple ``blurring'' of details 
from the spectral density function.  Without this ``smooth'' window, 
the loss of resolution would occur in a more complicated way. 

If we define the number of terms in the calculation: $N \equiv T/\Delta t$; 
and the discrete frequencies $f_k = (k-{\textstyle{1 \over 2}})/(2T), k=1,N$; 
then we can write the three dimensional digital transforms (following 
\ref{forwardtransform} and \ref{inversetransform}) as:
\begin{equation}
g_{AB} [f_k ] = \frac{{\left( {\Delta t} \right)^2 }}{T}\sum\limits_{j = 1}^N {\frac{{t_j }}{{f_k }}} s_{AB} [t_j ]D[t_j ]\sin [2\pi f_k t_j ], 
\end{equation}
for $k=1,N$, and conversely for $j=1,N$:
\begin{equation}
s_{AB} [t_j ]D[t_j ] = \frac{2}{{\Delta t}}\sum\limits_{j = 1}^N {\frac{{f_k }}{{t_j }}} g_{AB} [f_k ]\sin [2\pi f_k t_j ].
\end{equation}

To examine bias in the spectral density estimate, the discrete transform 
of \ref{sampop} can be taken.  Again if the lags are uniformly 
distributed over the sample volume, then to first order this can be 
written:
\begin{equation}
\langle g_{AB} [f]\rangle  = \gamma _{AB} [f] - \langle (\bar A - \mu _A )(\bar B - \mu _B )\rangle \sum\limits_{j = 1}^N {\frac{{t_j }}{f}\sin [2\pi ft_j ]} .
\end{equation}
The summation term in this equation is the discrete 
approximation to a delta function at the origin.  
That is, it is very large for $f \simeq 0$, but very small 
for other values of $f$.  The upshot of this is that 
the spectral density estimate is likely to be seriously biased 
due to uncertainty in the specimen averages only for $f \ge (1/T)$.  For
aspects of the estimates for which bias due to the finite 
sample is not a problem, the analysis of uncertainties is 
straightforward if $A$ and $B$ are assume to be Gaussian 
variates\cite{Blackman59}.  As mentioned above, one result 
of such analysis is that large uncertainties are associated with 
data on lags larger than half the sample size.  These 
uncertainties result in large uncertainties for all frequencies 
in the spectral density estimate.  Thus making the maximum lag 
allowed into the calculation ($T$) small compared to the sample 
size has two effects on the spectral density estimates:  It 
decreases spectral resolution but at the same time increases 
the stability of spectral density estimates for all frequencies.

\section{The statistics of sampling}

Suppose one measures the variable $A$ at $M$ points on a specimen, 
and wishes to determine an average value for $A$ in that specimen.  
The $M$ analysis ``points'' constitute the sample.  Standard 
practice is to calculate the sample mean from the measured values 
for $A$:
\begin{equation}
\bar A \equiv \frac{1}{M}\sum\limits_{i = 1}^M {A_i .} 
\end{equation}
This value $\bar A$ constitutes an ``unbiased'' estimate for the 
specimen (or ``population'') mean $\mu_A$ since the expected value for 
this measurement is indeed the population avearge, i.e.:
\begin{equation}
\langle \bar A\rangle  \equiv \mu _A .
\end{equation}

The object of this section is to provide a prediction for the 
error in this estimate.  To see how this error will depend on 
specimen heterogeneity, one need only consider the case in which all 
$M$ analysis points are clustered into a tiny volume much smaller than 
the grain size of a rock.  One would hardly expect to learn much 
about the composition of the whole rock from such an analysis, 
regardless of the size of $M$.  On the other hand, if the analysis 
points are randomly spread out over distances much larger than the 
average grain size, then for $M$ sufficiently large it should be 
possible to determine the rock composition arbitrarily well.  A nice 
feature of the approach adopted here is that the resulting answer 
will require no assumption about the random mixing of the constituent 
grains.  It instead follows simply from the definition of radial 
covariance.

The expected value for variance in the estimate of $\mu_A$ from 
measurements on the sample of $M$ points is:
\begin{equation}
\langle (\bar A - \mu _A )^2 \rangle  \equiv \frac{1}{{M^2 }}\sum\limits_{i = 1}^M {\sum\limits_{j = 1}^M {\langle (A_i  - \mu _A )(B_j  - \mu _B )\rangle } }
\label{varianceerror} 
\end{equation}
where the term in the summations is none other than $\sigma_{AA}[t_{ij}]$, i.e. 
the radial variance for separations $t_{ij}$ between ith and jth sample 
locations.  Thus the variance in estimates of the mean is just the average of 
radial auto-covariance for data from {\em pairs} of points in the sample.  In a 
similar fashion, an estimate of the covariance between sample estimates 
of $\mu_A$ and $\mu_B$ can be written as:
\begin{equation}
\langle (\bar A - \mu _A )(\bar B - \mu _B ) \rangle  \equiv \frac{1}{{M^2 }}\sum\limits_{i = 1}^M {\sum\limits_{j = 1}^M {\sigma_{AB}[t_{ij}]} } .
\label{covarianceerror}
\end{equation}
Thus knowledge of the radial covariance functions for a given specimen 
allows prediction of the uncertainty in estimates of means based on 
measurements from a limited sample.

The way in which sample size and compositional coarseness modifies estimates 
of uncertainty in the mean can be seen more clearly if the ``i=j'' terms are 
separated from the double sum in equation \ref{varianceerror}.  One can then 
write:
\begin{equation}
\langle (\bar A - \mu _A )^2 \rangle  = \{ 1 + (M - 1)\rho \} \frac{{\sigma _A ^2 }}{M} = \{ \begin{array}{*{20}c}
   {\frac{1}{M}{\sigma _A ^2 }} & {,\left| \rho  \right| \ll  \frac{1}{M}}  \\
   {} & {}  \\
   {\rho \sigma _A ^2 } & {,\left| \rho  \right| \gg  \frac{1}{M}}  \\
\end{array}
\label{errormean}
\end{equation}
where a {\em sample bias coefficient} for the sample configuration has 
been defined as an average over all pairs of sample points by:
\begin{equation}
\rho  \equiv \frac{1}{{M(M - 1)}}\sum\limits_{i = 1}^M {\sum\limits_{j \ne 1}^M {\frac{{\sigma _{AA} [t_{ij} ]}}{{\sigma _{AA} [0]}}.} } 
\label{samplebias}
\end{equation}
This sample bias coefficient is to first order dependent only 
upon the lag dependence of the covariance function for $A$, and on the 
sample size.  It, like the radial covariance function, is expected 
to approach zero for sample locations which are widely separated.
Hence, for sufficiently large sample regions, equation \ref{errormean} predicts that 
sample analysis points can be considered independent of one another.  
By the same token, however, for a sample of fixed size, an increase 
in the number of analysis points beyond $M \simeq \left|1/\rho \right|$ buys 
little decrease in uncertainty. 

If one wishes to estimate $\sigma_A^2$ from the same limited-sample 
measurement, the increase in uncertainty due to a sample which is too 
localized is even more severe.  If one measures the sample variance:
\begin{equation}
s_A ^2  \equiv \frac{1}{M}\sum\limits_{i = 1}^M {(A_i  - \bar A)^2 } ,
\end{equation}
then an unbiased estimate for $\sigma_A^2$ is obtained from the 
relation:
\begin{equation}
\sigma _A ^2  = \left( {\frac{M}{{M - 1}}} \right)\frac{{\langle s_A ^2 \rangle }}{{1 - \rho }} \cong \{ \begin{array}{*{20}c}
   {{\textstyle{M \over {M - 1}}}\langle s_A ^2 \rangle } & {,\left| \rho  \right| \ll {\textstyle{1 \over M}}}  \\
   {} & {}  \\
   {{\textstyle{1 \over {1 - \rho }}}\langle s_A ^2 \rangle } & {,\left| \rho  \right| \gg {\textstyle{1 \over M}}}  \\
\end{array}
\end{equation}
Equation \ref{errormean} can then be written as:
\begin{equation}
\langle (\bar A - \mu _A )^2 \rangle  = \frac{{1 + (M - 1)\rho }}{{(M - 1)(1 - \rho )}}\langle s_A ^2 \rangle , 
\end{equation}
which in the limits becomes
\begin{equation}
\langle (\bar A - \mu _A )^2 \rangle  \cong \{ \begin{array}{*{20}c}
   {{\textstyle{1 \over {M - 1}}}\langle s_A ^2 \rangle } & {,\left| \rho  \right| \ll {\textstyle{1 \over M}}}  \\
   {} & {}  \\
   {{\textstyle{\rho  \over {1 - \rho }}}\langle s_A ^2 \rangle } & {,\left| \rho  \right| \gg {\textstyle{1 \over M}}}  \\
\end{array}
\label{errormeansample}
\end{equation}
Again, these values equal those for independent sample locations 
if $\left| \rho \right|$ is sufficiently small.  However for a 
sufficiently small sampled {\em region}, $\rho$ approaches one and 
even hundreds of analysis points on such a small sample cannot 
result in reasonable uncertainties for our estimate of $\mu_A$.

For the case when the sampled region is not a set of discrete 
points but a continuous sampled volume, then two 
simplifications ensue.  First, the number of analysis points 
$M$ effectively goes to infinity, so that the limiting cases of 
equations \ref{errormean} and \ref{errormeansample} apply for which 
$\left| \rho \right| \gg \frac{1}{M}$.  Secondly, equation 
\ref{samplebias} can be rewritten:
\begin{equation}
\rho  = {\textstyle{{4\pi } \over V}}\int\limits_0^\infty  {\frac{{\sigma _{AA} [t]}}{{\sigma _{AA} [0]}}\xi [t]t^2 dt = } {\textstyle{{4\pi } \over V}}\int\limits_0^\infty  {\frac{{\gamma _{AA} [f]}}{{\sigma _{AA} [0]}}\Xi [f]f^2 df.}
\label{contsamplebias} 
\end{equation}
Here $\xi[t]$ is the containment probability defined in section \ref{wellmixed}, 
but this time 
with reference to the sample volume $V$ instead of a grain volume.  As 
before, $\Xi[f]$ is the radial Fourier transform of $\xi[t]$.  When viewed 
in this light, the quantity $\xi[t]/V$ is just the probability of two 
points in the sample being separated by distance $t$, per unit volume increment 
$4 \pi t^2 dt$.  The integration constraint \ref{volumeconstraint} becomes: 
\begin{equation}
4\pi \int\limits_0^\infty  {\frac{{\xi [t]}}{V}t^2 dt}  = \frac{{\Xi [0]}}{V} = 1.
\end{equation}
Because of the presence of $V$ in the denominator of equation \ref{contsamplebias}, 
a sample bias coefficient can also be calculated for two-dimensional and 
one-dimensional sample arrays.

For the purpose of this study, the most welcome facet of equation \ref{contsamplebias} 
is that the cube approximation of \ref{cubicapprox} can be utilized for both 
$\xi[t]$ and for $\sigma_{AA}[t]/\sigma_{AA}[0]$, in one case with reference to 
the sample volume and in the second case with reference to the specimen grain size, 
to provide a good general purpose estimate of sample bias coefficient as a 
function of sample size $D$ and grain size $d$:
\begin{equation}
\rho \left[ {\frac{D}{d}} \right] \cong \{ \begin{array}{*{20}c}
   {1 - \left( {\frac{D}{d}} \right) + \frac{2}{7}\left( {\frac{D}{d}} \right)^2 } & {,D \le d}.  \\
   {} & {}  \\
   {\left( {\frac{d}{D}} \right)^3  - \left( {\frac{d}{D}} \right)^4  + \frac{2}{7}\left( {\frac{d}{D}} \right)^5 } & {,D > d}.  \\
\end{array}
\end{equation}

\begin{acknowledgments}
Thanks for inspiration to Kahlil Gibran (1883-1931), because 
in this work we attempt to put into quantitative perspective 
his assertion that indeed ``I discovered the secret of the sea, 
in meditation upon the dew drop''.
\end{acknowledgments}
\bibliography{temrefs2.bib}

\appendix
\section{Heterogeneity of interplanetary dust}

One of the qualitatively striking features of material in 
chondritic interplanetary dust is the small, 
submicron size scale for chemical heterogeneity\cite{Fraundorf81}.  
This is manifest in three ways:  i) in the small (e.g. 10-100 nm) 
size of individual mineral grains;  ii) in the fact that compositions 
at locations separated by distances as small as 200 nm often seem 
to bear no relationi to one another, and  iii) by the remarkable 
observation\cite{Brownlee78a} that, in spite of their small 
size, nanogram specimens which probably sample different 
parent bodies show abundance ratios to silicon for 10 major 
elements which on the average agree within 40\%.  

Chondritic aggregates are not the ideal samples on which 
to begin making quantitative measurements of spatial heterogeneity, 
since contiguous sections of the aggregates available for 
such measurements are seldom more than several microns across.  
On the other hand, informative models of spatial heterogeneity 
are possible and further testing of those models may be 
possible with return of Stardust mission cometary dust 
specimens in 2006.  It is toward these ends that this appendix is 
designed.  In the second section of this appendix, it will be 
shown that the small grain size of the chondritic aggregates 
suggests that the 40\% spread in element to silicon ratios 
mentioned above is evidence for significant diversity in the 
compositions of chondritic aggregate source materials, in so 
far as those compositions remain intact in our samples.  In 
addition, comparisons with material in meteorites are discussed.  
In this section, the theoretical framework underlying the 
basic approach is addressed.  

  In discussing chondritic aggregate ``source materials'', it is 
important to address the fundamental problem in simple terms at 
the outset.  By ``source material'' we refer to the parent mass of a 
particular aggregate.  Implicit in this is the assumption that the 
aggregate was at one time part of a larger body.  The parent mass 
might, for example, be a geological zone in a comet.  If comet exteriors 
are largely undifferentiated, then the geological zone might be 
an accretionary layer.  In any case, it is obviously impossible 
to {\em deduce} the properties of a source material based on 
information which pertains only to a sample of that material.  On 
the other hand, it is reasonable to make {\em inferences} based on what 
is known, even if that is very little.  

To illustrate our predicament
with respect to chondritic aggregate materials, suppose someone 
was to provide you with a hand specimen of material, and without 
providing additional information, to ask for your best guess as to 
the nature of the source.  If the hand specimen was a crystal of 
uniform composition, you might guess that the parent material 
was a collection of similar crystals, in the absence of any hint 
as to other possible constituents.  But having had some experience 
with terrestrial rock formations, you might have strong doubts 
that the single crystal was a very representative sample.  If 
the hand specimen was instead an aggregate of thousands of millimeter 
sized crystals of three distinct minerals in apparently random 
juxtaposition with no obvious gradients or anistropy, you might have 
considerably more confidence in your estimate of source 
mineralogy, composition, and structure from the sample.  Of course, 
in either case the hand specimen might have been unrepresentative, 
but given a feel for the random way in which rocks are often 
put together, much more confidence would be accorded to inferences 
from the second type of sample.  One assumes that the parent material 
is a simple random mixture of crystals in the absence of 
evidence to the contrary.  In the remainder of this section, the 
implications of this assumption will be made quantitative with 
help from results from the paper to which this appendix is attached.

The measure of point-to-point heterogeneity central to this 
argument can be expressed in two complimentary forms, as the 
radial covariance function $\sigma_{AB}[t]$ or its 3-D Fourier 
transform, the spectral density $\gamma_{AB}[f]$.  Here $A$ and 
$B$ are both composition variables in the specimen (e.g. volume 
fraction of olivine, mass fraction of silicon, etc.)  When 
$A$ and $B$ represent the same variable, we often prefer 
displaying spectral density as the decomposition of standard 
deviation per decade of frequency (or size-scale).    

When a variety of structures is present in a specimen, one of 
course formally should Fourier transform the whole spatial array, 
thus in effect adding complex Fourier coefficients from each structure.  
However, when one can treat the structures as essentially 
uncorrelated in position with respect to one another (i.e. as 
ideally random in distribution), it is convenient instead to 
simply "add Fourier intensities (i.e. squared amplitudes)" from 
each object type.  

Thus for a random mixture of cubic grains of side $d$ which 
have various values for $A$ and $B$, it follows from 
equation \ref{cubicapprox} in section \ref{wellmixed} that
these functions are approximately:
\begin{equation}
\sigma _{AB} [t] \cong \{ \begin{array}{*{20}c}
   {\sigma _{AB} [0]\left( {1 - \frac{t}{{bd}}} \right)^2} & {,t \le bd}  \\
   {} & {}  \\
   0 & {,t > bd}  \\
\end{array}
\end{equation}
and
\begin{equation}
\gamma _{AB} [f] \cong \frac{{2 + \cos [2\pi fbd] - 3\sin [2\pi fbd]}}{{2\pi ^3 f^4 bd}}\sigma _{AB} [0].
\end{equation}
Here $b \equiv \left(\frac{30}{4 \pi} \right)^{1/3}$, 
and $\sigma_{AB}[0]$ is the covariance of $A$ and $B$ in the specimen.  
Similar expressions follow for a random mixture of spherical 
grains from equation \ref{sphericalcase}.

\begin{figure}[tbp]
\includegraphics{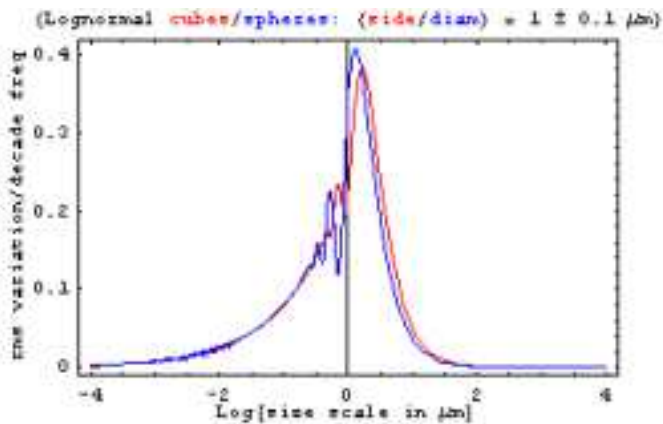}%
\caption{Root mean square compositional variability per decade of 
size-scale (or frequency) for log-normally distributed single-phase 
cubes one micron on a side, and spheres one micron in diameter.}
\label{Fig1}
\end{figure}

\begin{figure}[tbp]
\includegraphics{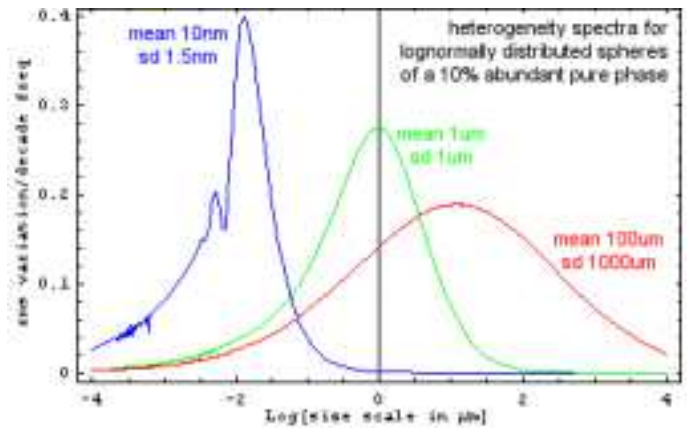}%
\caption{Root mean square compositional variability per decade of 
size-scale (or frequency) for log-normally distributed single-phase 
spheres with three different mean diameters and diameter standard deviations.}
\label{Fig2}
\end{figure}

For specimens composed of pure phase ($A=1$) grains, 
$10\%$ abundant by volume in 
an $A=0$ matrix, the resulting standard deviation 
decompositions for collections of lognormally distributed 
grains are illustrated in Figures \ref{Fig1} and \ref{Fig2}.
In particular, Figure \ref{Fig1} illustrates the 
effect of grain shape differences (cubes versus spheres) 
for a relatively tight distribution of grain sizes 
(size or diameter mean of $1 \mu m$ and standard deviation 
of $0.1 \mu m$.  Tighter size distributions (i.e. 
smaller ratios of standard deviation to mean size) result in 
even more oscillations in the high frequency (small size-scale) 
sides of the peak.

Figure \ref{Fig2} illustrates the effect of changing 
grain size and standard deviation.  Note that the 
effects of grain shape (e.g. Fourier ringing) are 
washed out as the range of sizes broadens.  We also 
expect this to be the case as individual grain symmetries 
decrease.  Of course, correlations in grain position or 
orientation might add new structure (e.g. lattice 
periodicities) to such heterogeneity spectra.

Our experience with fine-grained interplanetary dust particles
(typical sizes $8 \mu m$ in diameter) suggest that the 
porous or ``re-entrant'' material has a mean size for monomineralic 
grains (i.e. olivines, pyroxenes, sulfides) in the $0.1 \mu m$ size 
range\cite{Fraundorf81}.  However, particles with multi-micron sized monomineralic (e.g. olivine) grains with attached fine-grained 
material are also found.  Thus for example a well-mixed 
lognormal grain size 
distribution with mean near $0.2 \mu m$ but a much larger  
standard deviation may be appropriate for some types of 
interplanetary dust.  The well-mixed model could be tested on 
microtomed dust particles not available at the time of the 
original paper, as well as against fine-grained carbonaceous 
meteorite matrices which may prove to be compositionally 
much coarser.

The tools described in this paper further
allow one to model the relationship between individual dust 
particles and their ``parent bodies''.  For example, if the 
size and shape of individual grains is independent of composition, 
the correlation coefficient of a given particle to it's 
source might be written\cite{Fraundorf80}
\begin{equation}
\rho _{net}  \cong \sum\limits_{i = 1}^m {p_i \rho [\frac{D}{{d_i }}]} 
\end{equation}
Here $p_i$ is the volume fraction of grains of size $d_i$.
More careful work on this question is likely worthwhile, 
particularly in anticipation of the Stardust comet sample return mission 
in 2006, at which time multiple particles from a single 
identified cometary source will become available.  


\section{Log-log scale roughness spectroscopy}

Here we turn 
from studies of compositional heterogeneity in three 
dimensions to the study of height variations across 
a surface in two dimensions\cite{ruffspec93, ruffspec93b}.  
The transform equation (e.g. \ref{forwardtransform}) in two dimensions is:
\begin{equation}
\gamma _{AB} [f] = 2\pi \int\limits_0^\infty  t J_0 [2\pi f t] {C_{AB}[t] dt}, 
\label{2Dforward}
\end{equation}
where $J_0$ is the 0th order Bessel function of the first kind.  As with 
the inverse in three dimensions (equation \ref{inversetransform}), the reverse 
transform in two dimensions is the forward transform, with $t$ and $f$ exchanged.  
The corresponding $t=0$ relationship in 2D then becomes:
\begin{equation}
C_{AB} [0] \equiv \int\limits_0^\infty 2\pi {f} \gamma_{AB} df = \int\limits_{ - \infty }^\infty  {2\pi f^2 } \gamma_{AB} d(\ln f).
\label{t0case2D}
\end{equation}
Thus once again one might relate spectral density physically via a 
product (here $2 \pi f^2 \gamma_{AB}$) to 
that portion of the covariance $\sigma_{AB}[0]=C_{AB}[0]-\mu_A \mu_B$ between $A$ 
and $B$ in a given ``e-fold'' of frequency (or decade of frequency should one 
further multiply by $\ln 10$).

\begin{figure}[tbp]
\includegraphics{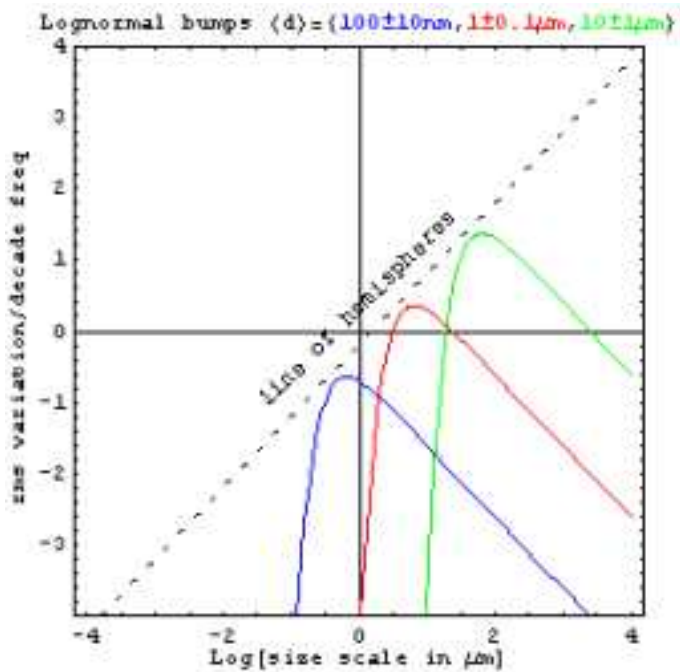}%
\caption{Root mean square roughness per decade of 
size-scale (or frequency) for log-normally distributed 
gaussian bumps three different mean diameters and 
diameter standard deviations.}
\label{Fig3}
\end{figure}

Consider for example a Gaussian bump, of half-width $d$ at 
half of its maximum height $A$.  If this bump is centered at 
the origin, it's height as a function of 
position is $h[t]=A \exp[-\frac{t^2}{2 d^2}]$.  
The 2D Fourier transform of this from \ref{2Dforward} is 
$H[f] = 2 \pi d^2 A \exp[-2 \pi^2 f^2 d^2]$.  Moving the 
bump from the origin will affect the transform's phase, 
but not the spectral density obtained by squaring 
it's amplitude: $\gamma_{hh}[f] = 4 \pi^2 d^4 A^2 \exp[-4 \pi^2 f^2 d^2 ]$.  
Inverse transforming this gives a height autocorrelation of 
$C_{hh}[t] = \pi d^2 A^2 \exp[- \frac{t^2}{4 d^2}]$, which 
becomes the autocovariance $\sigma_{hh}[t]$ as well if indeed the 
spatial region integrated over is large enough to render the average
height essentially zero.

Note in particular the ``line of hemispheres'' in Fig. \ref{Fig3}.  
Crossing above this basically signals 
the existence of objects on a given width scale which 
are taller than they are wide.  Grassy plains might 
represent an extreme in this regard.  At the 
other extreme (exceptional flatness), silicon 
wafers immediately after growth of an epitaxial 
layer, with $0.13 nm$ dimer row steps sometimes 
separated by a micron or more laterally represent 
the other extreme.  We anticipate providing some 
experimental examples of these extremes, and 
further application examples, in a 2nd 
revision.


\section{Lattice fringe covariance fingerprints}

In this appendix, covariances in high energy electron 
intensity as a function of scattering angle are 
examined as a tool for fingerprinting nanocrystal 
assemblages\cite{fringecov04}, given one or more high resolution lattice 
fringe images to work with.  Calculations of these for some 
common nanoparticle structures are underway, and an expansion 
of this discussion is anticipated in a 2nd revision as well.

\end{document}